\documentclass[PRL,twocolumn,showpacs,preprintnumbers,amsmath,amssymb,superscriptaddress]{revtex4}
\usepackage{graphicx}
\usepackage{epstopdf}
\usepackage{amssymb}
\usepackage{epstopdf}
\def\Ce {CeCoIn$_5 $ }

\begin{document}

\draft
\vspace{-11pt}

{\bf \noindent Comment on ``Texture in the Superconducting
Order Parameter of \Ce  Revealed by Nuclear Magnetic Resonance''}
\\

\vspace*{-0.4cm}
The study of the   
Fulde-Ferrell-Larkin-Ovchinnikov (FFLO) state \cite{FFLOdis}  has
been of considerable recent interest. Below the temperature $T^*$
which is believed to be the transition temperature ($T$) to the FFLO
phase in CeCoIn$_5$, K. Kakuyanagi  {\it et al.}
\cite{Kakuyanagi05} reported a composite  NMR spectrum with a tiny
component observed at frequencies corresponding to the normal
state signal. The results were interpreted as evidence for the
emergence of an FFLO state. 
This result is
inconsistent with two other NMR studies \cite{Mitrovic06,
Young06}. Mainly, the relative shift difference between the low
 $T$  superconducting (SC) and normal state reported in
\cite{Kakuyanagi05}  is much smaller than that  in
\cite{Mitrovic06, Young06}. Furthermore, the complex composite
lineshape observed close to $T^*$ is not seen in our
experiment. Finally, in the FFLO state we did not observe any signal
at frequencies corresponding to the normal state signal. In this
comment we show that the findings  in \mbox{Ref. \cite{Kakuyanagi05}}
do not reflect the true nature of the FFLO state but result from
excess RF excitation power used in that experiment.

\begin{figure}[t]
\begin{minipage}{0.98\hsize}
\centerline{\includegraphics[scale=0.43]{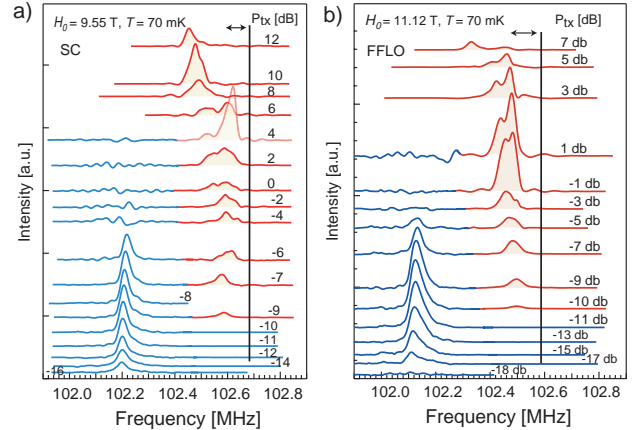}} 
\begin{minipage}{0.97\hsize}
\vspace*{-0.2cm} \caption[]{\label{Fig1}(Color online) \small
$^{115}$In(1) spectra  at  $H_0 || [100]$ obtained with  
the same pulse lengths but different 
   RF power.  The power (${\rm P_{tx}}$) is reported   relative to the   RF excitation field $H_1 \approx 2.2 \, {\rm mT}$ during the $\pi /2 = 3 \,{\rm \mu s}$ pulse, a value optimized in the normal state. The applied pulse power, neglecting 
the losses in the RF circuit,  at 0 dB is estimated to be less than 400 mW for $6 \,{\rm
\mu s}$ with a very  
low   duty cycle  (recordings  spaced by more than 8 s).    The colors are used to distinguish two parts of the same spectrum, where the  
  intensity of the red part (higher frequency)  is divided by  a factor of 10 compared to the blue ones. The black solid lines denote the corresponding frequency of the normal state spectra at $T_c$.  The arrows ($\leftrightarrow$) are the estimated shift reported in \cite{Kakuyanagi05}. 
{\bf a}) Spectra, $\langle -5/2 \leftrightarrow -7/2 \rangle$ transition, at 9.55 T     in the SC state. 
{\bf b})  Spectra, $\langle 1/2 \leftrightarrow 3/2 \rangle$ transition,  at 11.12 T  in the FFLO state. } 
\end{minipage}
\end{minipage}
\vspace*{-0.7cm}
\end{figure}

In  \mbox{Fig. \ref{Fig1}},   NMR spectra   at  $T \approx
70$ mK at \mbox{$H_0 = 9.55$ T}, in the SC state, and at $H_0 = 11.12$ T, in the  FFLO state,   are shown.  
The spectra were recorded using 
different RF  pulse powers while keeping all other parameters fixed.  The signals at the lower frequency (blue) part of the spectra, {\it i.e.} having low shift, 
are as those reported in  \cite{Mitrovic06}.  
The signals at higher frequencies  (in  red) are at relative shift comparable to those 
 in \cite{Kakuyanagi05}. 
Evidently, signals
with distinct shift  can be obtained depending on the excitation power.
The higher frequency  signal  
corresponds to the main line  in   \cite{Kakuyanagi05}. The
exact spectral shape and position of this line is dependent on
$\tau$, the time interval between the excitation and detection RF
pulses, and on the pulse power. This is in contrast to the signal at lower
frequencies. It is clear that the application of the  RF power in a certain range can lead to the double peak structure   solely in the FFLO state, as in  \cite{Kakuyanagi05}.  
Consequently, we deduce that the data reported in
\cite{Kakuyanagi05} do not reflect the true nature of neither the SC nor the FFLO
state.  
Moreover, in a singlet superconductor the absolute value of the shift  can only decrease with decreasing $T$. Thus, the lower shift  attests to the correctness of our results in  \cite{Mitrovic06}.

The most likely origin of the higher frequency signal is the RF heating  of the electronic system. This signal  is  detected at   frequencies comparable to those of the normal state   at $T \sim 1$ K. 
  Furthermore, only 
  this signal displays  some sample dependence. 
In fact, we found that this dependence  stems solely  from the exact size of the NMR coil, that is the power density  of the delivered RF. 
Lower frequency   signal obtained with low RF power  did not show any sample dependence.

As stated above, when the appropriate RF power is applied we did not observe any   NMR signal at low $T$
that corresponds to a normal state signal as in
\cite{Kakuyanagi05}. 
 This is not surprising, since a signal from
the spatial regions of vanishing order parameter should   not
necessarily appear at the normal state frequency \cite{Caroli64}.
Furthermore,  calculations  show that even if such a signal is to appear,  it should be located within the line corresponding to the vortex lattice lineshape  \cite{matchida}.  Obviously, this is   not what is observed  in \cite{Kakuyanagi05}. 
  Rather,  the weak  signal that   appears at the   frequency of the normal state in \cite{Kakuyanagi05} is well separated from the main NMR line in disagreement with  the theoretical predictions for the FFLO vortex state  \cite{matchida}.  
  
\noindent V. F. Mitrovi{\'c}$^{1}$, G. Koutroulakis$^{1}$, M.
Klanj{\v{s}}ek$^{2}$, M. Horvati{\'c}$^{2}$, C. Berthier$^{2}$, 
 G. Lapertot$^{3}$, J. Flouquet$^{3}$

\noindent
$^{1}$    Brown University, Providence, RI 02912, U.S.A.  \\
$^{2}$ GHMFL, CNRS,  
38042 Grenoble, France.  \\
$^{3}$ DRFMC, SPSMS, CEA Grenoble,   France.

\vspace{-0.4cm}
\bibliographystyle{apsrev}

\end{document}